# Orbital contributions in the element-resolved valence electronic structure of Bi$_2$Se$_3$

Cheng-Tai Kuo,[1,2,3*] Shih-Chieh Lin,[1,2] Jean-Pascal Rueff,[4,5] Zhesheng Chen,[4] Irene Aguilera,[6,†] Gustav Bihlmayer,[6] Lukasz Plucinski,[7] Ismael L. Graff,[8] Giuseppina Conti,[1,9] Ivan A. Vartanyants,[10,11] Claus M. Schneider,[1,7] Charles S. Fadley[1,2 ‡]

[1]*Department of Physics, University of California Davis, Davis, California 95616, USA*
[2]*Materials Sciences Division, Lawrence Berkeley National Laboratory, Berkeley, California 94720, USA*
[3]*Stanford Synchrotron Radiation Lightsource, SLAC National Accelerator Laboratory, Menlo Park, California 94025, USA*
[4]*Synchrotron SOLEIL, L'Orme des Merisiers, Saint-Aubin-BP48, 91192 Gif-sur-Yvette, France*
[5]*Sorbonne Université, CNRS, Laboratoire de Chimie Physique-Matière et Rayonnement, F-75005 Paris, France*
[6]*Peter Grünberg Institut and Institute for Advanced Simulation, Forschungszentrum Jülich and JARA, D-52425 Jülich, Germany*
[7]*Peter-Grünberg-Institut PGI-6, Forschungszentrum Jülich, Jülich 52425, Germany*
[8]*Department of Physics, Federal University of Paraná, Curitiba, Brazil*
[9]*Advanced Light Source, Lawrence Berkeley National Laboratory, Berkeley, California 94720, USA*
[10]*Deutsches Elektronen-Synchrotron DESY, Notkestraße 85, D-22607 Hamburg, Germany*
[11]*National Research Nuclear University MEPhI (Moscow Engineering Physics Institute), Kashirskoe Shosse 31, Moscow, 115409, Russia*
*ctkuo@slac.stanford.edu
[†]Present address: IEK5-Photovoltaik, Forschungszentrum Jülich, 52425 Jülich, Germany.
[‡]Deceased August 1, 2019.

## ABSTRACT

In this work, we studied the bulk band structure of a topological insulator (TI) Bi$_2$Se$_3$ and determined the contributions of the Bi and Se orbital states to the valence bands using standing wave-excited hard X-ray photoemission spectroscopy (SW-HAXPES). This SW technique can provide the element-resolved information and extract individual Bi and Se contributions to the Bi$_2$Se$_3$ valence band. Comparisons with density functional theory (DFT) calculations (LDA and *GW*) reveal that the Bi 6*s*, Bi 6*p*, and Se 4*p* states are dominant in the Bi$_2$Se$_3$ HAXPES valence band. These findings pave a way for studying the element-resolved band structure and orbital contributions of this class of TIs.

I. INTRODUCTION

In the last few years, topological insulators (TI) have been extensively studied due to their unique physical properties and great potential for device applications [1,2]. $Bi_2Se_3$ is a prototype material of a three-dimensional (3D) TI with a bulk band gap (~0.3 eV) and a gapless surface state consisting of a single Dirac cone, which has been discovered both theoretically and experimentally [3,4]. Prior to density functional theory (DFT) calculations, models starting from atomic energy levels and considering the crystal-field splitting and spin-orbit coupling successfully predicted band inversion and the existence of a single Dirac cone within a band gap [4,5]. Thereby, however, a simplified picture has been used by assuming that the *p* orbitals of Bi and Se atoms mainly contribute to the electronic behavior near the Fermi surface, thus neglecting the contributions of *s* orbitals. The neglect of *s* orbitals might be acceptable for the prediction of topological states. However, it is important to have a more quantitative information regarding the contributions of individual atomic orbital states to the valence states, because it helps to understand the bulk band structure of $Bi_2Se_3$, as well the nature of its band gap and band gap energy. Recently, more efforts in terms of DFT calculations have been done to study the total and partial densities of states (DOSs) in $Bi_2Se_3$ [6,7,8]. Compared to the theoretical studies, less experimental approaches have been utilized in this aspect. An experimental determination of partial DOSs with both element- and orbital-projected information in $Bi_2Se_3$ and in the long term for the whole class of 3D TIs is needed.

Numerous studies using synchrotron-based angle-resolved photoemission spectroscopy (ARPES) with excitation energies of 20-150 eV have been employed to study the electronic properties of this family of 3D TIs [1]. Low-energy ARPES is one of the key experimental techniques for investigating the electronic structure of TIs and its high energy resolution shows strength in resolving the topological and electronic states in the valence band (VB) [9,10,11]. However, the short inelastic mean free path of photoelectrons limits the probing depth, adding an extra challenge to differentiate the bulk and surface electronic structure. Nonetheless, Bianchi *et al*. tuned the $k_z$ of $Bi_2Se_3$ by varying the incidence photon energies and successfully differentiated the bulk and surface electronic states near the Fermi level [12]. Some ARPES experiments were carried out by using a laser source of 6~7 eV [13,14] or Xe lamp of 8.4 eV [15,16,17,18], which can enhance depth sensitivity, but the purpose was to improve the intensity and k-resolution for studying the physics of Dirac cone, which is beyond the scope of our work. Hard X-ray photoemission spectroscopy (HAXPES) extends the probing depth and is more sensitive to the bulk electronic structure. It is therefore useful for investigating the VB spectra of materials [19,20]. The drawbacks of HAXPES are relatively low energy resolution (~hundred meV) and phonon effects, which may limit its potential for resolving fine features in the VB and the k-resolved electronic structure [19,21,22]. Another advantage of HAXPES is the capability of generating a standing wave (SW) electric field within the desirable material by matching the length scale of the natural crystallographic atomic planes. SW-HAXPES collect photoelectrons modulated by the SW field from different depths in the c-axis, and it is a unique technique to determine the spatial distribution of composition and the element- or layer- resolved DOSs within a unit cell of a material, which has been applied in numerous systems [23,24,25,26,27,28,29,30,31].

In this work, we employed SW-HAXPES in studying $Bi_2Se_3$ and successfully observed the SW effects in both core-level and valence photoelectrons, providing the element-resolved information. Additionally, we discuss the $Bi_2Se_3$ bulk electronic structure in terms of the element-resolved valence spectra (the Bi- and Se-projected matrix-element weighted DOSs) directly in comparison to DFT calculations based on an all-electron method and subsequent *GW* calculations. Our results reveal the contributions of atomic orbitals (e.g. Bi 6*s*,*p* states) and attribute the quantity of each individual state to the $Bi_2Se_3$ HAXPES VB. Furthermore, the understanding of the $Bi_2Se_3$ band structure will be beneficial to both experimental and theoretical studies of the emergent properties in this class of TI.

## II. RESULTS AND DISCUSSION

$Bi_2Se_3$ has a rhombohedral crystal structure with space group ($R\bar{3}m$) [32], as shown in Figure 1(a). The material has a layered structure with five atomic layers forming one unit cell, known as a quintuple layer [3]. Each quintuple layer consists of two equivalent Se atomic layers (denoted as Se1), two equivalent Bi atomic layers (denoted as Bi), and a third Se atomic layer (denoted as Se2). The quintuple layers are held by weak van-der-Waals bonds. The Bi, Se1, and Se2 layers are well separated along the vertical axis, allowing the SW technique to derive site-specific information. Figure 1(b) shows the cross-section view geometry of the SW pattern generated in a $Bi_2Se_3$ crystal. A SW with its iso-intensity planes parallel to the diffracting planes is created by the interference between the incident and diffracted waves while satisfying the Bragg condition [33,34]. The phase difference between the incident and diffracted wave fields changes by π when the incidence angle moves from below to above the Bragg angle, thus scanning the SW by half of the wavelength with respect to the diffracting planes. Depending on the vertical locations of these atomic layers, the incidence-angle dependence of the core-level photoelectron intensity yields will show distinct modulations as to both shape and magnitude.

The SW-HAXPES measurements were performed at the beamline GALAXIES of SOLEIL. The details of this beamline can be found elsewhere [35,36]. The measurement was carried out at an excitation energy of 3000 eV and at room temperature. At this energy, the total experimental resolution is estimated to be about 300 meV. The $Bi_2Se_3$ crystal (HQ graphene) was *in situ* cleaved prior to the SW-HAXPES measurements. The survey spectrum shows no indication of the C 1$s$ and O 1$s$ surface contaminants. In this work, we have chosen the (006) instead of the (003) reflection as the vertical locations of Bi, Se1, and Se2 layers lead to the cancellation of SW modulation for (003) reflection. Figure 2 shows the SW-HAXPES results for core-level spectra and intensity yields as a function of incidence angle. The photoelectron spectra of Se 2$p_{3/2}$ and Bi 4$d$ at an off-Bragg angle are shown in Figs. 2(a) and (b). In the Se 2$p_{3/2}$ spectrum (Fig. 2(a)), only one component centered at 1422.4 eV is observed. We were expecting to observe two different chemical states for Se1 and Se2 since they are located at different positions. However, the observation of a single component in the Se 2$p_{3/2}$ spectrum indicates that the chemical states of Se1 and Se2 layers cannot be discriminated experimentally. As a consequence, we attribute the Se 2$p_{3/2}$ photoemission yield to both Se1 and Se2 layers. In the Bi 4$d$ spectrum (Fig. 2(b)), the spin-orbit splitting of bulk Bi 4$d_{5/2}$ (440.7 eV) and 4$d_{3/2}$ (464.6 eV) peaks are dominant (green line); in addition, we observed two additional features (blue and magenta line). The first feature (blue line) at 447.8 and 471.4 eV arises from plasmon-phonon coupling and the second feature (magenta line) at 458.3 and 481.8 eV arises from plasmon losses, both of which have been discovered and assigned by Biswas *et al.* [37]. As these three components originate from the bulk crystal, their photoemission yields show no difference.

Figure 2(c) and (d) show the SW photoemission yields of Se $2p_{3/2}$ and Bi $4d$ core levels (open circle) with the incidence angle ranging from 25.3° and 26.3°. These intensity scans are normalized to 1 at off-Bragg positions and have been simulated by SW theory (curve) [31,34]. The SW modeling shows good agreement with experiments. We observed that the Se $2p_{3/2}$ and Bi $4d$ photoemission yields for (006) reflection are out-of-phase; the Se $2p_{3/2}$ yield shows a maximum and the Bi $4d$ yield shows a minimum in intensity at ~25.8°. The yield intensity modulation of Bi $4d$ is around 7%. The modulation in intensity for the Se $2p_{3/2}$ yield is relatively small (~3%) due to the slight phase cancellation between the Se1 and Se2 layers. This observation of out-of-phase core-level photoemission yields allows the forthcoming extraction of individual Bi and Se photoemission cross-section-weighted partial DOSs from the $Bi_2Se_3$ valence band, which will be discussed later. Note that since the contributions from the Se1 and Se2 layers are indistinguishable in the Se $2p$ core level, we can only extract partial DOSs from the sum of the Se1 and Se2 layers.

To resolve the Bi and Se contributions to the $Bi_2Se_3$ VB, we use the following procedure. Under the influence of a SW electric field intensity, the $Bi_2Se_3$ VB yield $I_{VB}(E_B, \theta_{inc})$ depends on the core-level yields ($I_{Se\ 2p}(\theta_{inc})$ or $I_{Bi\ 4d}(\theta_{inc})$) and photoemission cross-section-weighted partial DOSs (i.e. $D_{Se}(E_B)$), as given by the following equation [23-31]:

$$I_{VB}(E_B, \theta_{inc}) = D_{Se}(E_B) I_{Se\ 2p}(\theta_{inc}) + D_{Bi}(E_B) I_{Bi\ 4d}(\theta_{inc}). \tag{1}$$

The core-level yields are normalized to 1 away from the Bragg reflection and adjusted in amplitude according to the differential cross-section ratios $\frac{d\sigma_{Se\,4s,4p}}{d\Omega} \Big/ \frac{d\sigma_{Se\,2p}}{d\Omega}$ and $\frac{d\sigma_{Bi\,6s,6p}}{d\Omega} \Big/ \frac{d\sigma_{Bi\,4d}}{d\Omega}$. The photoemission cross-section-weighted partial DOSs in Eq. (1) are related to the partial densities of states $\rho(E_B)$ and differential photoelectric cross sections $d\sigma/d\Omega$ of the dominant orbitals [38]. For example, $D_{Se}(E_B) = \sum_{i=Se\,4s,4p} \frac{d\sigma_i}{d\Omega} \rho_i(E_B)$. By using a least-square fitting method to each curve in $I_{VB}(E_B, \theta_{inc})$ at a $E_B$ with $I_{Se2p}(\theta_{inc})$ and $I_{Bi4d}(\theta_{inc})$ as basis functions, we have determined the elemental contributions $D_{Se\,or\,Bi}^{expt}(E_B)$ to the valence band $D_{Bi_2Se_3}^{expt}(E_B)$ (or abbreviated as $D_{Bi_2Se_3}^{expt}$). As mentioned earlier, this methodology in X-ray SW photoemission has been proved valid and successfully applied to many systems [23-31], but not yet to a 3D TI, Bi$_2$Se$_3$, which is unique to this study. It is important to mention that Eq. (1) reduces to the following equation when the SW field is absent; this is consistent with a generic form of photoemission cross-section-weighted DOSs:

$$D_{Bi_2Se_3}(E_B) = D_{Se}(E_B) + D_{Bi}(E_B) = \sum_{i=Se\,4s,4p} \frac{d\sigma_i}{d\Omega} \rho_i(E_B) + \sum_{j=Bi\,6s,6p} \frac{d\sigma_j}{d\Omega} \rho_j(E_B). \quad (2)$$

The experimental Se- and Bi-resolved DOSs, that are labeled as $D_{Se}^{expt}$ and $D_{Bi}^{expt}$, are used to compare with the partial DOSs obtained from the DFT calculations after taking photoemission cross section into consideration, distinguishing the individual contributions. The DFT and *GW* calculations were carried out using the full-potential linearized augmented planewave (FLAPW) method as implemented in the FLEUR code [39] and the SPEX code [40], respectively. The local density approximation (LDA) [41] was employed for the exchange-correlation potential of DFT. The product of muffin-tin radii times plane wave cutoff energy was 9.7 and a 5×5×5 Monkhorst-Pack k-point mesh was used for the self-consistent calculations (400 k-points and the tetrahedron method for the DOS calculations). The converge parameters for the *GW* calculations are the same as those reported in Ref. [42] except that the k-point sampling employed here is denser (8×8×8). The theoretical DOSs that are generated by more sophisticated *GW* calculations are demonstrated along with experiments and LDA results for comparison in Figures 3-5. We see that the theoretical DOSs by LDA and *GW* are practically unchanged, but the band gap of 0.34 eV is in better agreement with experimental data [43].

Figure 3(a) shows the experimental VB photoemission signal of Bi$_2$Se$_3$, $D_{Bi_2Se_3}^{expt}$, along with Bi- and Se-resolved contributions, $D_{Bi}^{expt}$ and $D_{Se}^{expt}$, which were derived from eq. (1). The theoretical photoemission cross-section-weighted total DOSs, and Se as well as Bi DOSs calculated by LDA and *GW* are demonstrated for comparison in Figure 3(b) and (c), respectively. The theoretical DOSs were calculated considering photoemission cross section ratios ($\sigma_{Bi\ 6p}/\sigma_{Bi\ 6s} \sim 1.91, \sigma_{Se\ 4s}/\sigma_{Bi\ 6s} \sim 1.68,$ and $\sigma_{Se\ 4p}/\sigma_{Bi\ 6s} \sim 1.86$) [44,45]. The theoretical results were shifted 0.4 eV with respect to higher binding energies to match the experimental results. The energy shift could be due to the electron doping induced by defects in the naturally grown Bi$_2$Se$_3$ crystal, which is quite usual [4]. We notice that the experimental DOSs generally have broader features in energy as compared to the theoretical DOSs - in spite of the high energy resolution - even though the center positions of these features have good agreements. In the experimental VB photoemission signal, two main features centered at ~1.5 and ~4 eV and a smaller feature centered at ~5.5 eV are observed. One can observe that the topological surface states are significantly suppressed due to the low cross section in HAXPES. In the experimental Se-resolved DOSs, there is one broad feature centered at ~1.7 eV; its shape shows good agreement with the theories. Three features centered at ~1.2, 3.8, and 5.5 eVs are found in the experimental Bi-resolved DOSs. The center positions of these features agree with the theoretical Bi-resolved DOSs.

However, two disagreements are observed between the experimental and theoretical DOSs. First, the peak at ~1.2 eV in the $D_{Bi}^{\text{expt}}$ is underestimated in $D_{Bi}^{\text{LDA}}$ and $D_{Bi}^{\text{GW}}$. Second, the relative integrated intensity of the experimental Bi-resolved DOSs is larger than the theoretical DOSs. The photoemission cross-section-weighted DOS $D(E_B)$ depends on the partial densities of states $\rho(E_B)$ derived from DFT calculations and differential photoelectric cross sections $d\sigma/d\Omega$, as mentioned earlier. The discrepancies between the experimental and theoretical curves may be related to either or both of them. It is known that the LDA and $GW$ results of Bi$_2$Se$_3$ show fairly good agreements with the ARPES results [42,43]. Additionally, a few prior studies reported that the cross sections for valence electrons of other solid state materials are not known quite well [25-28,46,47], therefore the tabulated photoemission cross sections of the atomic orbital states may be the main cause of the discrepancies. The theoretical DOSs of Bi $s$,$p$ states ($\rho_{Bi6s,6p}$) and Se $s$,$p$ ($\rho_{Se4s,4p}$) states are shown in Fig. 4. Fitting these theoretical orbital states to the experimental element-resolved DOSs, one can obtain the corrected values of photoemission cross sections, solving the mentioned two disagreements. We find that the relative cross section needs to be multiplied by an additional factor of 2 for theoretical Se 4$s$ and Bi 6$s$ DOSs in order to improve agreements in terms of the shape of Se- and Bi-resolved DOSs. The second disagreement of underestimated intensity of theoretical Bi-resolved DOSs can be corrected using the integrated intensity ratio $D_{Se}^{\text{expt}} / D_{Bi}^{\text{expt}} \sim 1.4$. These types of corrections have previously been found in SW-HAXPES studies of various transition metal oxide systems for accurate interpretations of orbital contributions to HAXPES VB yields [25,27,28]. Additionally, the relatively large values of the *s* or *p* level cross sections in the HAXPES VB spectra have widely been found in various systems [19-28,46,47] and the cause can be attributed to the deviation between solid state and atomic cross sections. Figure 5 compares the resulting corrected theoretical element-resolved and total DOSs with the experimental results. The corrected results show good agreements with the experimental ones.

The resulting corrected theoretical orbital contributions to $D_{Se}^{c\text{-}LDA}$ (or $D_{Se}^{c\text{-}GW}$) and $D_{Bi}^{c\text{-}LDA}$ (or $D_{Bi}^{c\text{-}GW}$) have been compiled in Figure 5(a) and (b), respectively, along with $D_{Se}^{expt}$ and $D_{Bi}^{expt}$ (curve for LDA and dot for *GW*). The corrected LDA and corrected *GW* are labeled as c-LDA and c-*GW*, respectively. The Se 4*p* states dominate $D_{Se}^{expt}$ and its main feature peaked at ~1.7 eV agrees well with the experimental results. On the contrary, Se 4*s* contributions are negligible in $D_{Se}^{expt}$. In addition, the experimental $D_{Bi}^{expt}$ consists of Bi 6*s* and 6*p* states, as shown in Fig. 5(b). The Bi 6*p* states contribute to the strong feature at the region of 2-6 eV, while Bi 6*s* states contribute to other features at region of ~0.5-1.5 eV. Figure 5(c) and (d) shows a comparison of atomic orbital contributions from Bi 6*s*, Bi 6*p*, Se 4*s*, and Se 4*p* states in the Bi$_2$Se$_3$ HAXPES VB by c-LDA and c-*GW*, respectively. The Se 4*p* and Bi 6*p* states are main components in the valence band and the intensity of Bi 6*s* states is relatively weak, but the states are located very close to the valence band maximum (VBM) edge of Bi$_2$Se$_3$. Although the Bi 6*p* states are dominant in $D_{Bi}^{expt}$, the contributions of the Bi 6*s* states are still significant, especially in a relatively narrow and pronounced feature (0.5-1.5 eV), which is located in the proximity of the VBM, Dirac cone, and Fermi level. As discussed in the beginning, despite a successful prediction of the existence of topological surface states and the insulating energy gap of $Bi_2Se_3$, these prior works assume that the *p* orbitals of Bi and Se atoms mainly contribute to the electronic behavior near the Fermi surface and neglect the contributions of the *s* orbitals [3-8]. Our experimental SW-HAXPES results clearly show that the contributions of Bi 6*s* states have been underestimated so far regardless of calculation methods. Since both DFT results (LDA and *GW*) show no significant difference in $\rho_{Bi6s}$, one can conclude that the underestimation of Bi 6*s* states might mainly originate from the deviation between solid state and atomic cross sections, as discussed earlier.

To have a quantitative estimate of the composition of these states near the VBM edge, one can integrate their DOS intensities along binding energies (0 ~1.2 eV) and then normalize them by the sum of total integrated intensities. The small amount of theoretical DOSs intensity at binding energies between 0 and 0.3 eV for all the states is not taken into consideration, because these theoretical DOSs stem from the conduction band minimum (CBM) edge. Using the c-LDA results as an example, in the $Bi_2Se_3$ VBM edge, it consists of 64.9% Se 4*p*, 22.9% Bi 6*s*, 10.6% Bi 6*p*, and 1.6% Se 4*s* states. Interestingly, the composition of Bi 6*s* states is even higher than that of Bi 6*p* states near the VBM. The electronic states in the region between 1 eV and the Fermi level can be attributed to the bulk and surface band structure [37], surface topological states [4,48], the quantity of orbital-selective spin texture [13]. Therefore, our findings suggest that it is important to consider the contributions of *s*-orbital states in the electronic properties for the future studies. These findings should help the future theoretical calculations of $Bi_2Se_3$ band structures and the interpretation of the photoelectron valence band.

## III. CONCLUSION

In summary, we have used SW-HAXPES to study the bulk band structure of a 3D TI, $Bi_2Se_3$. By SW-HAXPES, we observed the Se $2p_{3/2}$ and Bi $4d$ photoemission yields are out-of-phase, allowing us to extract individual Bi and Se photoemission cross-section-weighted DOSs from a photoelectron $Bi_2Se_3$ valence band yield. Upon the comparison between experimental and DFT results (LDA and *GW*), we find that Se $4p$, Bi $6s$, and Bi $6p$ states are dominant in the $Bi_2Se_3$ VBM edge after accurately determining solid state cross sections. A pronounced feature at the region of 0.5-1.5 eV, which is in the proximity of the VBM and Fermi level, is attributed to the contributions of Bi $6s$ states. These findings are beneficial to the future studies of emergent properties of $Bi_2Se_3$ both experimentally and theoretically. As demonstrated, Bragg-reflection SW photoemission is thus promising for studying $Bi_2Se_3$ and the class of similar 3D TI, such as $Bi_2Te_3$ and $Sb_2Te_3$.

**ACKNOWLEDGMENTS**

We thank Simon Moser for sharing his experience concerning $Bi_2Se_3$ sample preparation. Charles S. Fadley unfortunately deceased on August 1, 2019; we are grateful for his significant contributions to this work. We thank synchrotron SOLEIL (via Proposal No. 20171604 and 99180118) for access to Beamline GALAXIES that contributed to the results presented here. This work was supported by DOE Contract No. DE-SC0014697 through the University of California Davis (salary for C.-T.K, S.-C.L. and C.S.F.). C.S.F. has also been supported for salary by the Director, Office of Science, Office of Basic Energy Sciences (BSE), Materials Sciences and Engineering (MSE) Division, of the U.S. Department of Energy under Contract No. DE-AC02-05CH11231, through the Laboratory Directed Research and Development Program of Lawrence Berkeley National Laboratory, through a DOE BES MSE grant at the University of California Davis from the X-ray Scattering Program under Contract DE-SC0014697, through the APTCOM Project, "Laboratoire d'Excellence Physics Atom Light Matter" (LabEx PALM) overseen by the French National Research Agency (ANR) as part of the "Investissements d'Avenir" program, and from the Jűlich Research Center, Peter Grűnberg Institute, PGI-6. Use of the Stanford Synchrotron Radiation Lightsource, SLAC National Accelerator Laboratory, is supported by the U.S. Department of Energy, Office of Science, Office of Basic Energy Sciences under Contract No. DE-AC02-76SF00515. I. L. G. wishes to thank Brazilian scientific agencies CNPQ (Project No. 200789/2017-1) and CAPES (CAPES-PrInt-UFPR) for their financial support.

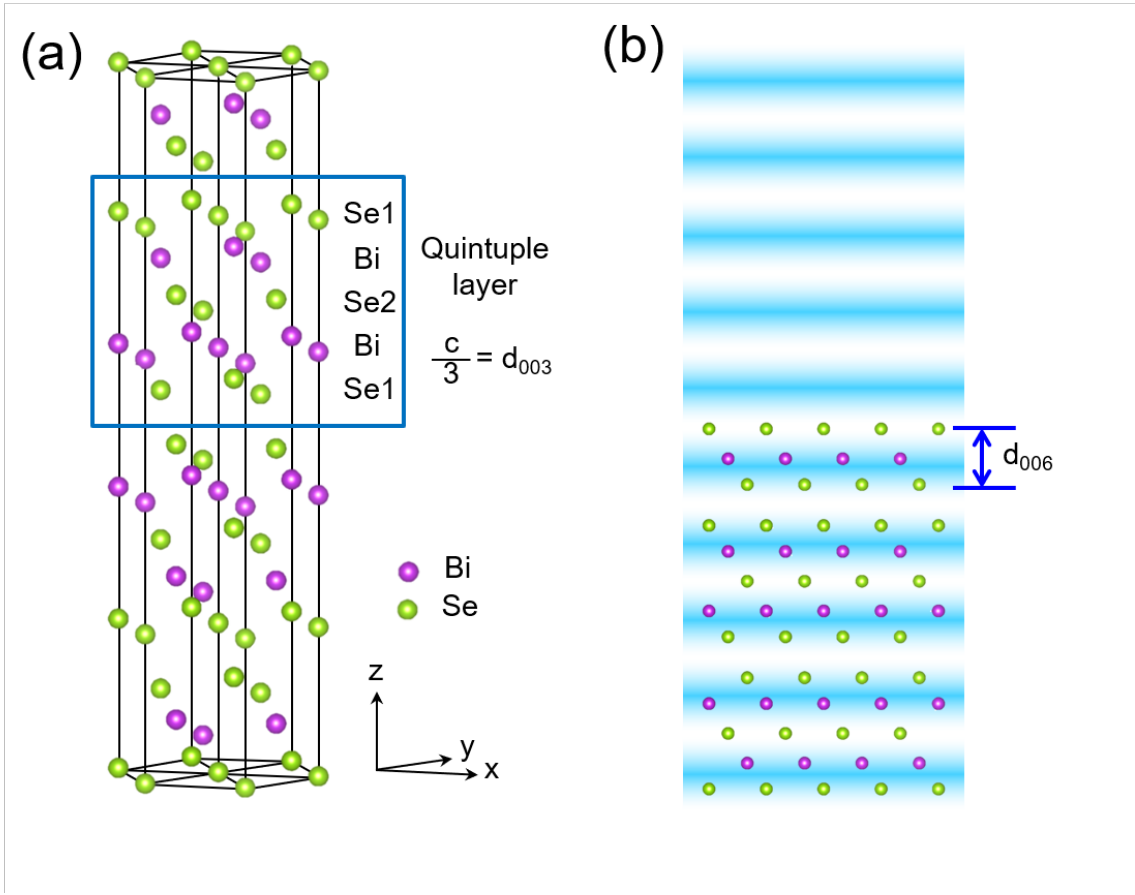

FIG. 1. (a) Rhombohedral crystal structure of $Bi_2Se_3$. A quintuple layer with Se1-Bi-Se2-Bi-Se1 is indicated by the blue rectangle. (b) Side view of the standing wave (SW) generated by the $Bi_2Se_3$ (006) reflection and different atomic species are excited with different intensity of SW electric field depending on the position of the SW nodes and antinodes.

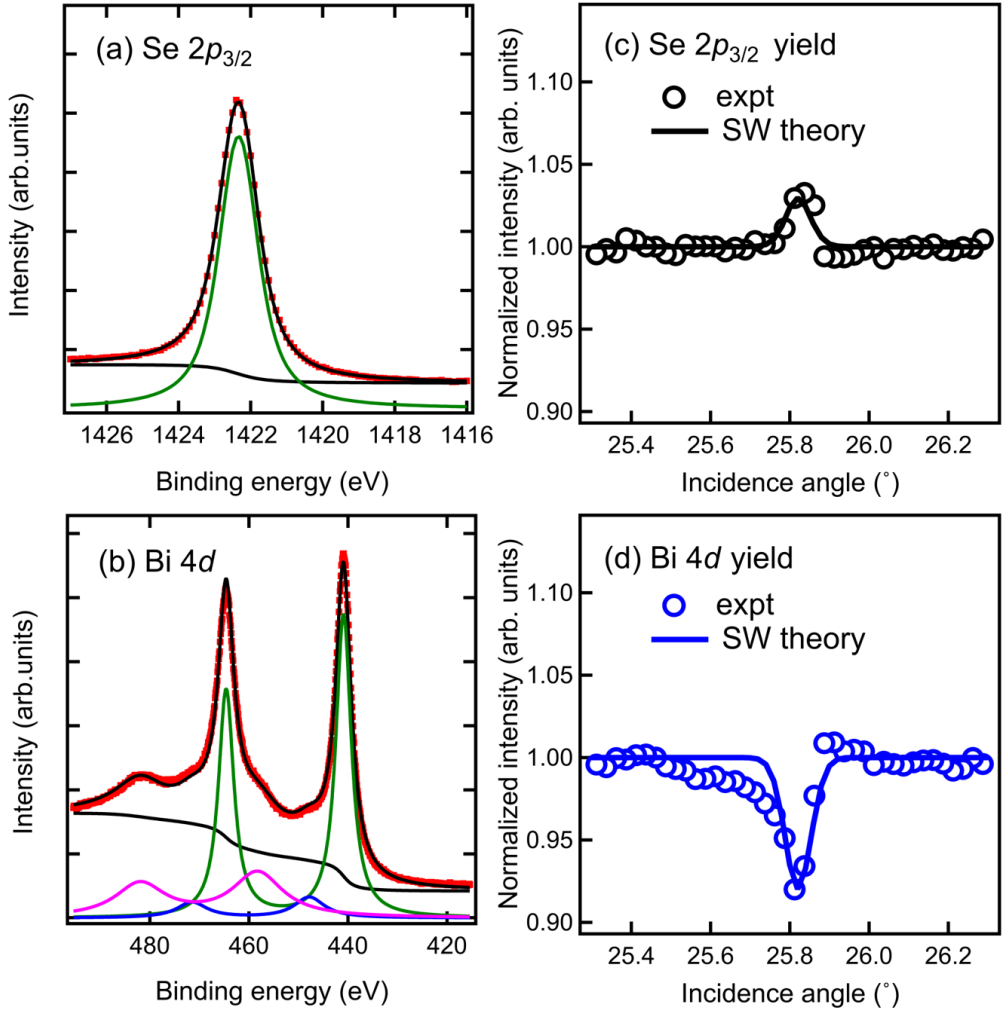

FIG. 2. SW-HAXPES core-level spectra and yields from the Bi$_2$Se$_3$ at $h\nu$ = 3000 eV. Core-level spectra of (a) Se $2p_{3/2}$ and (b) Bi $4d$ at an off-Bragg incidence angle. The core-level peak intensities are derived by fitting with a Voigt line shape (in green, blue, and magenta) and a Shirley background (in black). The corresponding experimental photoemission intensity yields of Se $2p_{3/2}$ and Bi $4d$ are plotted in (c) and (d), respectively, along with the SW theory simulations.

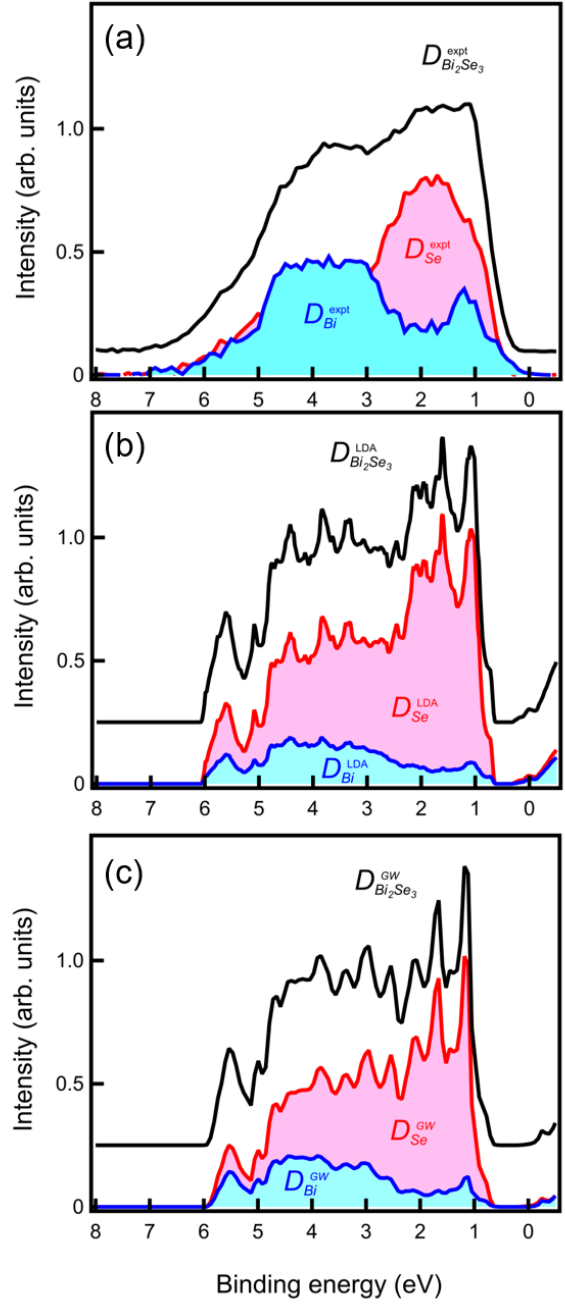

FIG. 3. (a) Experimental HAXPES valence band of $Bi_2Se_3$, $D_{Bi_2Se_3}(E_B)$, together with Se-resolved DOSs, $D_{Se}^{expt}(E_B)$, and Bi-resolved DOSs, $D_{Bi}^{expt}(E_B)$. Photoemission cross-section-weighted total DOSs calculated from LDA and *GW* are shown in (b) and (c), respectively, for comparison.

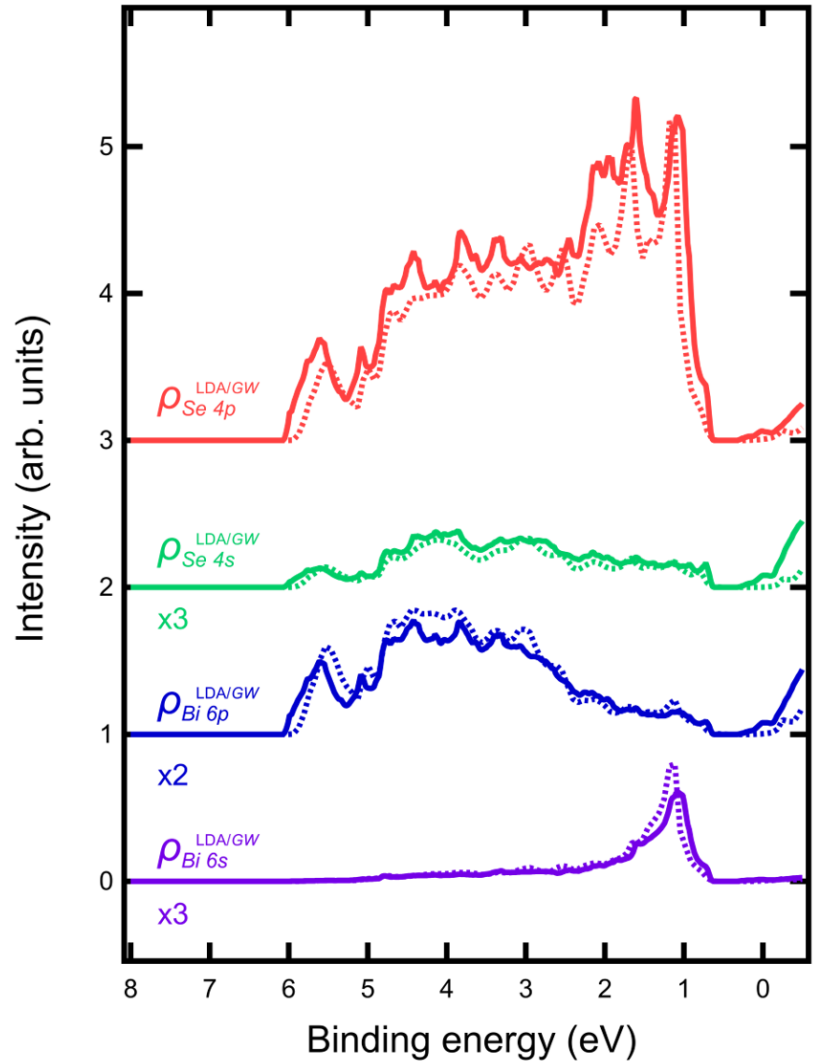

FIG. 4. Theoretical atomic orbital contributions to the $Bi_2Se_3$ valence band. The LDA results are color curves and $GW$ results are color dots.

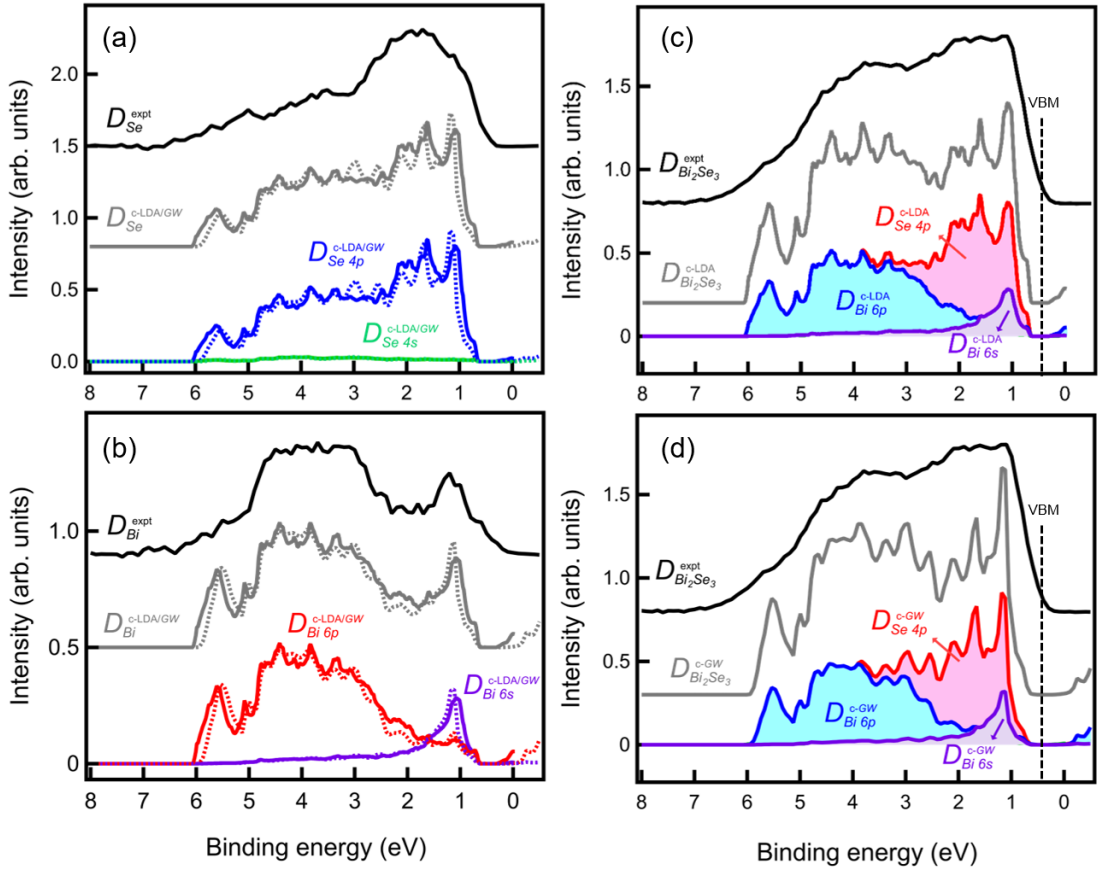

FIG. 5. (a) Comparison of experimental Se-resolved DOS, $D_{Se}^{\text{expt}}(E_B)$, and corrected theoretical orbital contributions to $D_{Se}^{\text{c-LDA}}(E_B)$ and $D_{Se}^{\text{c-}GW}(E_B)$. (b) Comparison of experimental Bi-resolved DOS, $D_{Bi}^{\text{expt}}(E_B)$, and corrected theoretical orbital contributions to $D_{Bi}^{\text{c-LDA}}(E_B)$ and $D_{Bi}^{\text{c-}GW}(E_B)$. The corrected LDA and corrected $GW$ are labeled as c-LDA and c-$GW$, respectively. In (a) and (b), the c-LDA results are color curves and c-$GW$ results are color dots. (c) and (d) show comparisons of experimental VB $D_{Bi_2Se_3}^{\text{expt}}(E_B)$ and c-LDA as well as c-$GW$ results.